\documentclass[12pt]{iopart}
\usepackage{graphics}
\usepackage{epsfig}

\def\apj{{\sl Astrophys.\ J. \ }}
\def\apjl{{\sl Astrophys.\ J.\ Lett. \ }}
\def\apjs{{\sl Astrophys.\ J.\ Supp. \ }}

\def\astroph#1{{\tt astro-ph/#1}}
\def\grqc#1{{\tt gr-qc/#1}}
\def\hepth#1{{\tt hep-th/#1}}

\def\ijmpd{{\sl International\ J.\ Mod.\ Phys. \ D \ }}

\def\lrr{{\sl Liv.\ Rev. Rel. \ }}
\def\mnras{{\sl MNRAS \ }}
\def\n{{\sl Nature \ }}

\def\np{{\sl Nucl.\ Phys. \ }}
\def\plb{{\sl Phys.\ Lett.\ B \ }}

\def\pr{{\sl Phys.\ Rep. \ }}
\def\prd{{\sl Phys.\ Rev.\ D \ }}

\def\prl{{\sl Phys.\ Rev.\ Lett. \ }}

\def\rmp{{\sl Rev.\ Mod.\ Phys. \ }}

\newcommand{\gsim}{\,\lower2truept\hbox{${>\atop\hbox{\raise4truept\hbox{$\sim$}}}$}\,}
\newcommand{\pp}{~~~.}
\newcommand{\vv}{~~~,}

\def\etal{{\rm et al.$\,$}}

\newcommand{\be}{\begin{equation}}
\newcommand{\ee}{\end{equation}}
\newcommand{\bea}{\begin{eqnarray}}
\newcommand{\eea}{\end{eqnarray}}

\begin{document}

\title[Scaling solutions in scalar-tensor cosmologies]{Scaling solutions in scalar-tensor cosmologies}

\author{Valeria Pettorino$^{1,2}$, Carlo Baccigalupi$^{3,4,5}$, Francesca Perrotta$^{3,4,5}$}
\address{$^{1}$ Universit\`{a} di Napoli {\it Federico II} and INFN, Sezione di
Napoli, \\ Complesso Universitario di Monte Sant'Angelo, Via
Cintia, I-80126 Napoli, Italy \\ $^{2}$ Universit\`{a} di Torino,
Dipartimento di fisica teorica, Via Giuria 1, Torino, Italy \\
$^{3}$ Institute f$\ddot{\rm u}$r Theoretische Astrophisik,
Universit$\ddot{\rm a}$t Heidelberg, Albert-berle Str. 2, D-69120
Heidelberg, Germany
\\ $^{4}$ SISSA/ISAS, Via Beirut 4, I-34014 Trieste, Italy \\ $^{5}$ INFN,
Sezione di Trieste, Via Valerio 2, I-34127 Trieste, Italy}

\begin{abstract}
The possibility of a connection between dark energy and gravity
through a direct coupling in the Lagrangian of the underlying
theory has acquired an increasing interest due to the recently
discovered capability of the extended quintessence model to
encompass the fine-tuning problem of the cosmological constant.
The gravity induced ``R-boost'' mechanism is indeed responsible
for an early, enhanced scalar field dynamics, by virtue of which
the residual imprint of a wide set of initial field values is
cancelled out. The initial conditions problem is particularly
relevant, as the most recent observations indicate that the Dark
Energy equation of state approaches, at the present time, the
cosmological constant value, $w_{DE}=-1$; if confirmed, such
observational evidence would cancel the advantage of a standard,
minimally coupled scalar field as a dark energy candidate instead
of the cosmological constant, because of the huge fine tuning it
would require. We give here a general classification of the
scalar-tensor gravity theories admitting R-boost solutions scaling
as a power of the cosmological redshift, outlining those behaving
as an attractor for the quintessence field. In particular, we show
that all the R-boost solutions with the dark energy density
scaling as the relativistic matter or shallower represent
attractors. This analysis is exhaustive as for the classification
of the couplings which admit R-boost and the subsequent
enlargement of the basin of attraction enclosing the initial
scalar field values.
\end{abstract}

\maketitle

\section{Introduction}
\label{i} In the last few years, cosmology has gone through a deep
revision of the basic ideas on which it was used to rely. In
particular, the most puzzling problem for the old, ``standard''
CDM scenario, arised when the observations of distant Ia
Supernovae evidenced an accelerated expansion of the Universe,
through the reconstructed magnitude-redshift relations
\cite{riess_etal_1998,perlmutter_etal_1999}.  Since these very
early indications, the case for a ``Dark Energy" component able to
accelerate the cosmic expansion became increasingly stronger when
the observations of the CMB anisotropies, together with large
scale structure data, clearly revealed a very close-to-flat,
low-density Universe
\cite{bennett_etal_2003,tegmark_etal_2004}. \\
Many phenomenological explanations for the dark energy involve
classical, minimally-coupled scalar fields violating the strong
energy condition (``quintessence'' fields, \cite{Qin}-\cite{Qfin}
and references therein), or a phantom energy (see, e.g.,
\cite{GD}) which, as a cosmological constant, violates the weak
energy condition. Most of these models, however, suffer the
worrying problem of initial conditions, in the sense that an
accurate and sometimes unphysical tuning of the initial values of
a field is required in order to reproduce the cosmological
conditions observed today (in particular, the dark energy equation
of state $w_{DE}$ and the density parameter of the dark energy
component). This problem, affecting almost all dark energy models
involving scalar fields, is partially alleviated when, depending
on the potential on which the field is assumed to evolve, the
scalar field equation of motion admits tracking solutions: in that
case, indeed, the present value of the cosmological parameters, as
requested by the observations, can be reached starting from some
more or less extended range of initial values of the field.
However, even for those models admitting a tracking behavior, the
problem of initial values  is now becoming particularly serious,
because the observation bounds on the dark energy equation of
state are increasingly converging towards a value of $w_{DE}$ very
close to $-1$. It is known that, the closer $w_{DE}$ is to the
cosmological constant value, the smaller the range of allowed
initial value for the corresponding scalar field has to be, since
the dynamics of such a field is extremely constrained by the
flatness of the potential in which the field evolves
\cite{Bludman}. Furthermore, the best fit of the latest Sn Ia data
(the Gold dataset,
 \cite{Riessnew})  includes values $w_{DE}< -1$, which cannot
be obtained in the context of minimally-coupled quintessence
models. In light of these considerations, it is worth to point our
attention to extensions of General Relativity where, standing the
presence of a scalar field as a dark energy candidate, the
dynamics of the field itself can be, at early times, strongly
modified by gravitational effects. In particular, we will focus on
scalar-tensor theories (see, e.g., \cite{fujii_maeda_2003}) where
a scalar field, non-minimally coupled to Gravity, acts as a dark
energy component at recent times
\cite{sahni}-\cite{allemandi_etal_2004}. A dark energy component,
as it arises in scalar-tensor theories, has been proven to have
several advantages with respect to minimally-coupled scalar
fields: the direct coupling of the field to the Ricci scalar in
the Lagrangian of the model, makes the field undergoing an
enhanced dynamics at early times, known as ``R-boost", with the
appealing consequence that the characteristic thickness of the
basin of attraction for tracker solutions is preserved even for
$w_{DE}$ close to $-1$ \cite{matarrese_etal_2004}. Furthermore, as
noticed in \cite{boisseau_etal_2000} \cite{torres} \cite{Perivola}
\cite{mingxing_etal_2005}, such an ``extended quintessence'' (EQ)
component can cross the cosmological constant value, getting
$w_{DE}< -1 $. In the past however, those interesting results were
obtained considering particular cases of scalar-tensor theories of
gravity, while a general treatment is still missing. The purpose
of the present paper is to fill this gap. We provide here a
general classification of all the possible scalar-tensor models of
Dark Energy which admit R-boost trajectories for which the
quintessence energy density scales as a power law of the
cosmological scale factor. In the following, we refer to those as
scaling solutions. Although in a different context, for purpose
and methodology our paper follows the footsteps of
\cite{liddle_scherrer_1999}
performed in the context of minimally coupled quintessence models. \\
The plan of the paper is the following:  in section \ref{frame},
we describe the working framework, giving the basic definitions cosmological
equations; in section \ref{coc}, we search for generic scaling
solutions of the Klein-Gordon equation, and classify the coupling
functions according to the resulting type of scaling; in section
\ref{attractors} we check these solutions for stability, verifying
which ones among the theories we found produce stable attractors;
finally, in section \ref{conclusions}, we draw our conclusions.

\section{Framework in generalized theories of gravity}
\label{frame} The class of Generalized Theories of Gravity we refer to is described by the action \be
\label{action} S = \int d^4x \sqrt{- g}\left[ \frac{1}{2\kappa} f(\phi,R) - \frac{1}{2}
\omega(\phi)\phi^{;\mu}\phi_{;\mu} - V(\phi) + {\cal{L}}_{\rm{fluid}}\right]\vv \ee where $g$ is the determinant
of the background metric, $R$ is the Ricci scalar and $\phi$ is a scalar field whose kinetic energy and potential
are specified by $\omega(\phi)$ and $V(\phi)$, respectively. ${\cal{L}}_{\rm{fluid}}$ includes contributions from
all components different from $\phi$ and $\kappa=8\pi G_{*}$ represents the bare gravitational constant. We also
work in natural units, $c=1$. The action above has been first considered in a cosmological context in
\cite{hwang_1991}. In particular, the classes of theories in which $f(\phi,R)$ assumes the simple form
$f(\phi,R)/2k = F(\phi)R/2$ have been considered in the context of dark energy cosmologies (extended
quintessence, see \cite{matarrese_etal_2004} and references therein). Note that the gravitational constant $G_*$
differs from the one measured in Cavendish like experiments by corrections being negligible in the limit
$\omega_{JBD} \gg 1$ \cite{esposito-farese_polarski} where $\omega_{JBD}$ is defined as \be \omega_{JBD} \equiv
\frac{F}{{F_{\phi}}^2} \ee and $F_{\phi}$ is the derivative of $F$ with respect to $\phi$. For a review of the
allowed values of the $\omega_{JBD}$ parameter and other constraints on generalized theories of gravity we send
back to \cite{barrow_2004}- \cite{bertotti_etal_2003}.

Compared to general relativity, the Lagrangian has been
generalized by introducing an explicit coupling between the Ricci
scalar and the scalar field, achieved by replacing the usual
gravity term $R/16\pi G$ with the function $f(\phi,R)/2\kappa$.
This new term, which has the effect of introducing a spacetime
dependent gravitational constant, may either be interpreted as an
explicit coupling between the quintessence field $\phi$ and
gravity, or as a pure geometrical modification of general
relativity admitting a non-linear dependence on $R$.

In the assumption of flat cosmologies, the line element can be written as $ ds^2 = -\, dt^2 + a^2(t)
\delta_{ij}dx^i dx^j \,, $ where $a(t)$ is the scale factor and $t$ represents the cosmic time variable; the
expansion of the Universe and the dynamics of the field are determined by Friedmann and Klein Gordon equations
\be \label{frw} {H}^2 = {\left(\frac{\dot{a}}{a}\right)}^2 = \frac{1}{3F}\left(\rho_{fluid} +
\frac{1}{2}{\dot{\phi}}^2 + V(\phi) - 3{H}{\dot{F}}\right)\ , \ee \be \label{kg} \ddot{\phi} + 3{H}\dot{\phi} =
\frac{1}{2}F_{\phi}R - V_{\phi} \ , \ee where dot means derivative with respect to the cosmic time $t$,
$F_{\phi}$ and $V_{\phi}$ are the derivatives of the coupling $F$ and of the quintessence potential $V$ with
respect to $\phi$. In this context, the Ricci scalar can be written in terms of the cosmological content of the
Universe \be \label{ricci} R=-\frac{1}{F}\left[-\rho_{fluid} + 3p_{fluid} + \dot{\phi}^2 - 4V +
3\ddot{F}+9H\dot{F}\right]\vv \ee where $\rho_{fluid}$ and $p_{fluid}$ are the energy density and pressure summed
up over all possible cosmological components but $\phi$.

As it was shown in \cite{hwang_1991}, all species but $\phi$ satisfy the usual conservation equations
$\dot{\rho}_i =-3 H(\rho_i + p_i)$, where $\rho_i$ and $p_i$ are respectively the energy density and pressure of
the $i$-th component. Unlike minimal coupled models, in extended quintessence the energy density \be
\label{rho_notcons} \rho_{\phi} = \frac{1}{2} {\dot{\phi}}^2 + V(\phi) \ee is not conserved, i.e. it does not
obey the relation $\dot{\rho}_{\phi}+3H(\rho_{\phi} +p_{\phi})=0$; nevertheless, one can still define a conserved
expression for the energy density including the contributions from the non-minimal coupling
\cite{perrotta_baccigalupi_2002} which reads as \be \label{rho_cons} {\rho}_{\phi}^{cons.} = \frac{1}{2}
{\dot{\phi}}^2 + V(\phi) - 3 H \dot{F} +3H^{2}\left(\frac{1}{\kappa}-F\right)\ .\ee The latter expression can be
generally very different from (\ref{rho_notcons}), mostly because of the last term, proportional to the
cosmological critical density and active whenever the theory differs from general relativity. Although that
difference is small to match the observational constraints, the presence of $H^2$ makes it relevant. This may
have important consequences for the dynamics of the dark energy density perturbations, leading to effects like
gravitational dragging \cite{perrotta_baccigalupi_2002}. However, as we already stressed, in the present work we
look for trajectories of the scalar field as a function of time, determined by the effective gravitational
potential in the Klein Gordon equation. In the energy density, this contribution is entirely kinetic
\cite{baccigalupi_etal_2000}, which therefore is the relevant energy density component for our purpose, as we see
in detail in the next Section. In other words, we are interested in attractor paths which solve the Klein Gordon
equation for the field, which is unique regardless of the definition of the energy density. In this perspective,
it is convenient to define a third expression related to the dark energy density \cite{baccigalupi_etal_2000}:
\be \label{rho_dyn} \tilde{\rho}_{\phi} = \frac{1}{2} {\dot{\phi}}^2 + V(\phi) - 3 H \dot{F}\ .\ee The quantity
above has a direct intuitive meaning for the dynamics of cosmological expansion, as it regroups the scalar field
terms which compete with $\rho_{fluid}$ in the Friedmann equation (\ref{frw}).

The cases in which $F(\phi)$ has a quadratic form (both induced
gravity and non minimal coupling theories
\cite{perrotta_baccigalupi_etal,baccigalupi_etal_2000}) or an
exponential form \cite{pettorino_baccigalupi_etal_2005} have been
widely investigated and reached their main achievement in what
concerns the early dynamics of the dark energy field
\cite{matarrese_etal_2004}. It was shown that as soon as there are
non-relativistic species in the radiation dominated era, the new
term containing $R$ in the Klein Gordon equation has a diverging
behavior \be \label{ricci_sempl} R \simeq
\frac{1}{F}\frac{\rho_{mnr0}}{a^3} \vv \ee where $\rho_{mnr0}$ is
the present value of the energy density of the components which
are non-relativistic at the time in which the R-boost occurs,
representing the only non zero contribution of the term
$-\rho_{fluid}+3p_{fluid}$ in eq.(\ref{ricci}); the remaining
terms in (\ref{ricci}) are negligible for $a$ sufficiently small.
The R-boost is caused by the effective, time dependent potential
arising in the Klein Gordon equation because of the presence of
the non-minimal coupling. The main point to be stressed in view of
the following analysis is that, at least for the particular
choices of coupling investigated up to now, the R-boost guaranties
an attractor behavior. This is a crucial aspect for extended
quintessence because it provides a valid alternative to minimal
coupled tracker fields, achieving attractors by means of their
standard potential \cite{liddle_scherrer_1999}
\cite{tsujikawa_etal_2004}; as we stressed above, this property
may disappear if the present dark energy equation of state gets
close to -1 \cite{Bludman}. On the other hand, the R-boost remains
a viable mechanism to keep a large basin of attraction if
$w_{\phi}$ approaches $-1$ \cite{matarrese_etal_2004}. These
scenario have been recently constrained with the available cosmic
microwave background and large scale structure data
\cite{acquaviva_etal_2005}.

In this work we provide the general analysis required to classify
the scalar tensor theories of gravity which have attractor
solutions to the Klein Gordon equation because of the non-minimal
coupling with the Ricci scalar.
With this purpose in mind, we will operate in the framework described
by equations $(\ref{action}) - (\ref{ricci})$ but we will not fix a
specific expression for the coupling $F(\phi)$. On the contrary, we will
try to classify the possible forms of the coupling which give rise
to R-boost trajectories behaving as attractors in the early
Universe. In order to proceed in this direction, we will follow
closely what has been done in \cite{liddle_scherrer_1999} in
the case of minimal coupling, where the authors classify the
allowed forms of the true potential $V(\phi)$ in order to have
scaling solutions.

\section{Classification of the couplings}
\label{coc} Let's consider a cosmological model described by
$(\ref{action}) - (\ref{ricci})$ containing both a scalar field
$\phi$ coupled to gravity through the function $F(\phi)$ and a
contribution of perfect fluid components with pressure and energy
density generically indicated with $p_{fluid}$ and $\rho_{fluid}$.
Assuming that in the Friedmann equation the terms involving the
scalar field are negligible with respect to the fluid energy
density \be \label{rf_gg_rphi} \rho_{fluid} \gg
\tilde{\rho}_{\phi} \vv \ee which is true in the epoch of early
Universe we are interested in, and that the fluid scales as \be
\label{rhofluid} \rho_{fluid} \propto a^{-m} \vv \ee we want to
find the forms of the coupling $F(\phi)$ which admit solutions of
the equation (\ref{kg}) for which \be \label{rhophi}
\tilde{\rho}_{\phi}\propto a^{-n} \vv \ee where $m$ and $n$ are
two integer positive defined numbers.

Assuming that the eq.(\ref{ricci_sempl}) for the Ricci scalar is
still valid and since the potential $V(\phi)$ in eq.(\ref{kg}) has
no relevant effect up to recent times, we can rewrite the Klein
Gordon equation in the following way: \be \label{kg_sempl}
\ddot{\phi} + 3{H}\dot{\phi} =
\frac{1}{2}\frac{\rho_{mnr0}}{a^3}\frac{F_{\phi}}{F} \pp \ee

The Friedmann equation ($\ref{frw}$), assuming that the perfect fluid is the dominant component, allows us to
calculate the behavior with time of the scale factor $a(t)$ as \be \label{at} a =
\left(\frac{t}{t_{*}}\right)^{2/m}\vv \ee where $t_{*}$ stands for a fixed reference time in the RDE and we
neglected any initial condition, assuming to work with $t$ large enough with respect to $t_{*}$. In eq.(\ref{at})
we have assumed that the time dependence of $F$ in the Friedmann equation is modest enough not to affect
significantly the dependence on the scale factor of the right hand side in the Friedmann equation (\ref{frw}).
Also, the last two terms in the Friedmann equation must be small in order to satisfy the condition
($\ref{rf_gg_rphi}$). However, we will still verify a posteriori that these assumptions are plausible.
Substituting in eq.(\ref{kg_sempl}) we get: \be \label{kg_t} \ddot{\phi} = -\frac{6}{m}\frac{1}{t}\dot{\phi} +
\frac{1}{2}\rho_{mnr0}\left(\frac{t_\ast}{t} \right)^{6/m}\frac{F_{\phi}}{F} \pp \ee At this stage the energy
density of the field $\phi$ is mainly given by its kinetic contribution, acquired through slow rolling onto the
effective gravitational potential in the Klein Gordon equation. Thus $\tilde{\rho}_{\phi}\simeq \dot{\phi}^2/2$
and assuming the desired scaling behavior (\ref{rhophi}) we obtain $\phi$ time dependence \be \label{phidot_t}
\dot{\phi} \propto t^{-{n/m}} \pp \ee We will now proceed by distinguishing the two cases $m=n$ or $m \neq n$.

\subsection{Case $m=n$}\label{mn}
If $m=n$ then $\dot{\phi} \propto t^{-1}$ and integrating this expression we get \be \label{phi_meqn} \phi = A
\ln{\frac{t}{t_*}}+\phi_* \ee where $A$ is a constant with the dimensions of a field, i.e. proportional to the
Planck mass $m_P=1/\sqrt(G)$ in our units. Substituting (\ref{phi_meqn}) and its first and second derivatives in
eq.(\ref{kg_t}) we obtain the following expression \be \label{ffphi} \frac{F_{\phi}}{F} = \frac{2A}{\rho_{mnr0}}
\frac{1}{{t_*}^2} \left(\frac{6}{m}-1 \right) e^{\left[{2 \left( \frac{3}{m} -1 \right)
\frac{\phi-\phi_*}{A}}\right]} \pp \ee If $m \neq 3$, the condition on the coupling, obtained by integrating
eq.(\ref{ffphi}), is \be \label{f_meqn} F = F_{*} e^{\left[B \left(e^{C(\phi-\phi_*)}-1\right)\right]} \vv \ee
where \be B = \frac{A^2}{\rho_{mnr0}} \frac{1}{{t_*}^2} \frac{6-m}{3-m} \vv \,\,\,\,\, C =
\frac{2}{A}\left(\frac{3}{m}-1\right) \ee and $F_*$ is the value the coupling has at $t_*$. Note that the
combination $\rho_{mnr0}t_{*}^2$ has a direct interpretation in terms of the abundance of the non-relativistic
components at the $t_{*}$ time; indeed $\rho_{mnr0}t_{*}^2\propto (a_*/a_0)^3\rho_{mnr*}/H_{*}^{2}$.

If $m = 3$, which is the case of a Universe dominated by ordinary matter, eq.(\ref{ffphi}) becomes \be
\label{ffphi_meqn3} \frac{F_{\phi}}{F} = \frac{2A}{\rho_{mnr0}}\frac{1}{t_*^2} \ee and the form of the allowed
coupling is \be \label{f_meqn3} F = F_* e^{\left[\frac{2A}{\rho_{mnr0}}\frac{1}{t_*^2}(\phi - \phi_*)\right]} \pp
\ee As expected, this is right the form of the coupling chosen in \cite{pettorino_baccigalupi_etal_2005} if we
define $A \equiv \frac{\xi}{m_P}\frac{{t_*^2} \rho_{mnr0}}{{2}} $, where $\xi$ is an adimensional constant and
$m_P = 1/\sqrt{G}$ is the Planck mass in natural units. In the regime $\phi\gg\phi_{*}$, the coupling becomes
exactly the one exploited in \cite{pettorino_baccigalupi_etal_2005}.

Notice that in minimal coupling theories the case $m=n$ corresponds to a scenario in which the dark
energy scales as the dominant component, thus never achieving acceleration unless such regime is
broken by some physical mechanism, as in the case of quintessence with exponential potential
\cite{wetterich_1988,liddle_scherrer_1999}. On the other hand, in scalar-tensor theories of gravity
the case $m = n$ is fully exploitable in its context, since it has
been obtained precisely with the assumption (\ref{rf_gg_rphi}) and neglecting the true potential $V$ in
the Klein Gordon equation (\ref{kg}). Actually, its relevance is on the capability to provide an attractor
mechanism when the dark energy is sub-dominant, independently on the form of the potential energy driving
acceleration today. Indeed, it does not exclude a different behavior of the $\phi$ field at present time,
when the potential $V$ starts to have a dominant effect on the dynamics of $\phi$.

\subsection{Case $m \neq n$}\label{mnn}
Integrating eq.(\ref{phidot_t}) in the case $m \neq n$ we get: \be \label{phi_mdivn} \phi = \widetilde{A}
\left(t^{1-\frac{n}{m}}- {t_*}^{1-\frac{n}{m}}\right) +\phi_* \vv \ee where $\widetilde{A}$ has the dimensions of
a time derivative of a field. Substituting this expression in eq.(\ref{kg_t}), together with its first and second
derivatives, we obtain the following condition \be \label{ffphimdivn} \frac{F_{\phi}}{F} =
\frac{\widetilde{C}}{\widetilde{A}} {\left(\frac{\phi - \phi_*}{\widetilde{A}} + {t_*}^{1-\frac{n}{m}}
\right)}^{\widetilde{B}} \vv \ee where \be \label{C_mdivn} \widetilde{C} = \frac{2
{\widetilde{A}}^2}{\rho_{mnr0}}\frac{1}{{t_*}^{6/m}}\frac{(m-n)(6-n)}{m^2} \,\,\,\,\,\,\,\,\,\,\,\, \widetilde{B}
= \frac{6-m-n}{m-n} \pp \ee As in Section \ref{mn}, note the combination
$\rho_{mnr0}t_{*}^{6/m}\propto(a_*/a_0)^3\rho_{mnr*}/H_{*}^{6/m}$. Again we have to distinguish the two cases in
which the exponent $\widetilde{B}$ is equal or not to $-1$.

For $\widetilde{B} = -1$ the condition (\ref{ffphimdivn}) gives, after integration, polynomials of $\phi$ \be
\label{f_mdivnB1} F = F_* \left[1 + \frac{\phi -
\phi_*}{\widetilde{A}}{t_*}^{\frac{n}{m}-1}\right]^{\widetilde{C}} \pp \ee Notice that if the solution
(\ref{phi_mdivn}) is exactly a power law of the time $t$, namely $\widetilde{A}t_{*}^{1-\frac{n}{m}}$, the
coupling (\ref{f_mdivnB1}) becomes exactly a power law as well, as in induced gravity models (see e.g.
\cite{perrotta_baccigalupi_etal} and references therein). Therefore, this class of gravity theories admit scaling
solutions: however, those may arise in the radiation dominated era only, as $\widetilde{B}=-1$ with $m=4$ induces
$n=3$; on the other hand, the same condition with $m=3$ is not satisfied regardless of the value of $n$.

For $\widetilde{B} \neq -1$ the result of the integration of eq.(\ref{ffphimdivn}) is \be \label{f_mdivn} F = F_*
\,
e^{\left[\frac{\widetilde{C}}{1+\widetilde{B}}\left[\left(\frac{\phi-\phi_*}{\widetilde{A}}+{t_*}^{1-\frac{n}{m}}\right)
^{\widetilde{B}+1}-\left({t_*}^{1-\frac{n}{m}}\right)^{\widetilde{B}+1}\right]\right]} \pp \ee Notice that this
includes the exponential case when $\widetilde{B}=0$ and thus $m+n = 6$ with $m \neq n$. As a consequence, if
$m=4$ and $n=2$ we obtain again the coupling investigated in \cite{pettorino_baccigalupi_etal_2005} for the case
of radiation dominated era. The latter case may be obtained exactly in the limit in which $t_{*}$ and $\phi_{*}$
may be neglected in (\ref{phi_mdivn}). Note that expression (\ref{f_mdivn}) is generalized to values of
$\widetilde{B} \neq 0$ and includes exponential functions with general coupling constant $\widetilde{A}$
contained in $\widetilde{C}$ and defined in (\ref{phi_mdivn}). \\ The same case, $m=4=2n$ yielding
$\widetilde{B}=0$, also corresponds to the R-boost solution exploited in \cite{baccigalupi_etal_2000} in the
radiation dominated era. The reason why our formalism does not show that the form of the non-minimal coupling in
that case, $F=1/2\kappa+\xi\phi^2$, is compatible with such a solution, is the following. For small values of the
coupling constant $\xi$, the field dynamics is correspondingly reduced. Eventually one enters the regime in which
the second term in the right hand side of (\ref{phi_mdivn}) dominates over the first one, yielding a scalar field
value as a function of time which is effectively constant; this is a clearly transient phase, as eventually such
regime is broken, but before that the relation (\ref{ffphimdivn}) remains approximately true. Although this is
formally not a scaling solution as that found in the exponential case \cite{pettorino_baccigalupi_etal_2005}, the
values of $\xi$ may be chosen so small that for interesting initial conditions on $\phi$, its variation due to
the R-boost is not relevant at all relevant epochs in the radiation dominated era, keeping the solution $m=4=2n$
effectively valid also for the coupling considered in \cite{baccigalupi_etal_2000}. We may refer to the solutions
of the Friedmann and Klein Gordon equations in these scenarios as {\it transient} scaling solutions, meaning that
they hold until the true R-boost dynamics takes over moving the field value away from its initial condition. Note
finally that the same reasoning applies in general in all cases where $F$ is the sum of a constant plus a
positive power of the field, normalized by a coupling constant.


\subsection{Summary and consistency criteria}\label{sacc}
We have found, up to now, all possible choices of the coupling $F$ which can have scaling solutions verifying
(\ref{rf_gg_rphi}) - (\ref{rhophi}), in the sense that there are no other forms of the coupling admitting scaling
behavior in scalar-tensor theories of gravity; for some of them, like in the non-minimal coupling considered in
\cite{baccigalupi_etal_2000}, transient scaling solutions are admitted in the time interval preceeding the epoch
in which the R-boost motions moves the field away from its initial condition. We have found, as expected,
exponential forms of the coupling \cite{pettorino_baccigalupi_etal_2005} which have, however, been generalized to
other choices of $m$, $n$ and the $\widetilde{B}$ exponent in (\ref{f_mdivn}); also, we have seen that
(\ref{f_mdivnB1}) allows polynomials of $\phi$. Eq.(\ref{f_meqn}) also suggests that a new family for the
coupling $F$ might be allowed, in the case $m=n$, namely made by exponentials of exponentials. Though no other
coupling can allow for scaling solutions, we have by now no guarantee that the general solution of all these
expressions for the coupling will indeed
have an attractor behavior and this is what we are going to investigate in the next section. \\
Before moving to that, let us briefly check the consistency of the scaling solutions we found with the
assumptions we made. They are essentially three, concerning equations (\ref{frw}), (\ref{ricci}) and (\ref{at}).
It is important to note that the variation of $F$ induces corrections which are small in the limit
$\omega_{JBD}\gg 1$. For example, $\dot{F}/F=\dot{\phi}F_{\phi}/F\propto\dot{\phi}/\sqrt{\omega_{JBD}}$;
therefore, even if the field dynamics may be important as in (\ref{phidot_t}), the coupling constant may be
chosen small in order to yield a small variation in time of $F$. More precisely, it is easy to see that the
kinetic contribution in (\ref{frw}) is the lowest order term in $1/\omega_{JBD}$; all the others, involving a
change in $F$, yield terms like $F_{\phi}\dot{\phi}$, which are of higher order due to the presence of
$F_{\phi}$, which brings another $1/sqrt{\omega_{JBD}}$. Indeed, it is important to stress again that the R-boost
dynamics is caused primarily by a non-zero Ricci scalar, diverging in the early universe if at least one
non-relativistic species is present, and not only to the underlying scalar-tensor gravity theory. Thus in the
limit of a small coupling, all the three approximations mentioned above are satisfied. However, it is interesting
to push the analysis a little further here, by computing the scaling of the terms in (\ref{frw},\ref{ricci})
coming from the scalar-tensor coupling. Concerning the last term in (\ref{frw}), which may be written as
$-HF_{\phi}\dot{\phi}/F$, using (\ref{kg_t}) and (\ref{phidot_t}) it may be easily verified that it scales as
$1/t^{\frac{2m+2n-6}{m}}$; when compared with the first term in the right hand side of (\ref{frw}), it yields the
condition $n<3$ both in matter and radiation dominated eras. Note that if one ignores the issue related with the
coupling strength mentioned above, this relation is quite stringent, confining all the scaling solution in
scalar-tensor theories of gravity to possess a shape not steeper than $a^{-3}$. The same criteria lead to $n<3$
and $n<2$ in the matter and radiation dominated eras, respectively, for neglecting the last two terms in the
right hand side of (\ref{ricci}). Also in this case, this requirement may be bypassed by working in a small
scalar-tensor coupling regime.

We now turn to study which scaling solutions represent attractors for the field dynamics.

\section{Attractors}
\label{attractors} Our purpose here is to check whether the various forms of the coupling $F$ found in the
previous section lead to attractor solutions to the Klein Gordon equation (\ref{kg_sempl}). With this aim, we
will investigate whether the particular solutions found for the cases $m=n$ and $m \neq n$ are indeed
attractors or not. Following the criteria developed in \cite{liddle_scherrer_1999}, we will proceed by
linearizing the Klein Gordon equation with small exponential perturbations $\e^{\lambda \tau}$ around the
critical point, represented, as we will se, by our particular solution. At a linear level, the attractor
behavior will be guaranteed whenever the perturbation will converge to zero with time. We shall also
investigate numerically a few cases without
linearization. \\
With the following change of variable \be \label{u_def} u =
\frac{\phi}{\phi_e} \ee where $\phi_e$ is the exact particular
solution of eq.(\ref{kg_t}),  eq.(\ref{kg_sempl}) can be rewritten
as \be \label{kg_u} \ddot{u} \phi_e + \dot{u} \left[2\dot{\phi_e}
+ 3H\phi_e \right] =
\frac{\rho_{mnr0}}{2a^3}\frac{F_{\phi}}{F}\left(1-u \right) \pp
\ee We will now distinguish what happens in the two cases
discussed in the previous section, for $m \neq n$ and $m = n$.

\subsection{Attractor behavior for $m \neq n$}
As we have discussed in the previous section, in the case $m \neq n$ the coupling needs to satisfy
eq.(\ref{ffphimdivn}) and the exact solution $\phi_e$ depends on time as in eq.(\ref{phi_mdivn}). For simplicity
in the following we will consider $t \gg t_*$ in such a way that the $t_*$ term in eq.(\ref{phi_mdivn}) can be
neglected; also, for $t$ large enough, the initial condition $\phi_*$ can be neglected too and
eq.(\ref{phi_mdivn}) reduces to $\phi = \widetilde{A} \, t^{1-\frac{n}{m}}$, where $\widetilde{A}$ is an
arbitrary constant as defined in eq.(\ref{phi_mdivn}). Substituting the first and second derivatives of $\phi_e$
in eq.(\ref{kg_sempl}) we get \be \label{ffphi_mdn_e}
\frac{\rho_{mnr0}}{2a^3}\frac{F_{\phi}}{F}=\frac{(6-n)(m-n)}{m^2} \frac{\phi_e}{t^2} \pp \ee Using this
expression in eq.(\ref{kg_u}) and with the change of variable \be \label{tau_def} \tau = \ln \frac{t}{t_\ast} \vv
\ee we obtain \be \label{kg__umdivn} u'' + \frac{m-2n+6}{m} u' = \frac{(6-n)(m-n)}{m^2}(1-u) \vv \ee where the
derivatives are calculated with respect to $\tau$. As we can see, this equation admits a critical point for $u =
1$ and $u' = 0$. If we consider a perturbation $\delta u$ around the critical point, such that $u = 1 + \delta u$
we can rewrite eq.(\ref{kg__umdivn}) as \be \label{kg__umdivn_pert} \delta u'' + \frac{m-2n+6}{m} \delta u' = -
\frac{(6-n)(m-n)}{m^2}\delta u \ee which is a homogeneous differential equation of the second order in $\tau$
with constant coefficients. The generic expression of the perturbation $\delta u$ is thus a combination of
exponential terms ($\delta u = c_1 e^{\lambda_1 \tau} + c_2 e^{\lambda_2 \tau}$) where $c_1$ and $c_2$ are
arbitrary constants and $\lambda_{1,2}$ are equal to \be \label{l_mdivn} \lambda_{1,2} = -\frac{m-2n+6}{2m} \pm
\frac{\left|m-6\right|}{2m} \pp \ee It is easy to see that the eigenvalues are real and negative for $n <
\,$min$\{6,m\}$. As a consequence, for these values of the parameters, the perturbation will go to zero with time
and the solution of the Klein Gordon equation (\ref{kg__umdivn}) will converge to $u = 1$ making $\phi = \phi_e$
to behave as a stable attractor. Notice that this discussion includes the case of the exponential coupling
investigated in \cite{pettorino_baccigalupi_etal_2005}.


\begin{figure}
\begin{center}
   \includegraphics[width=7.5cm]{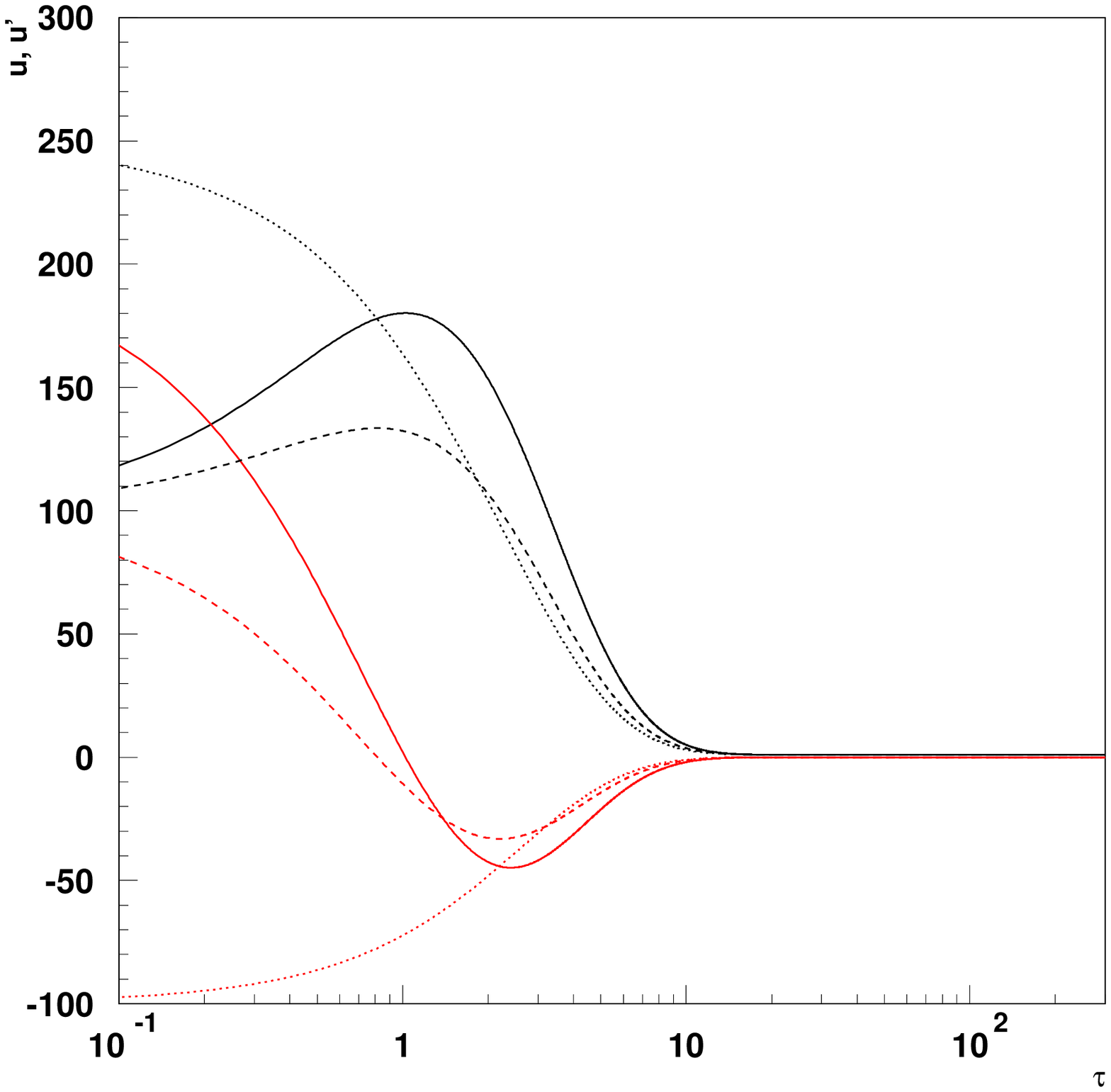}
   \includegraphics[width=7.5cm]{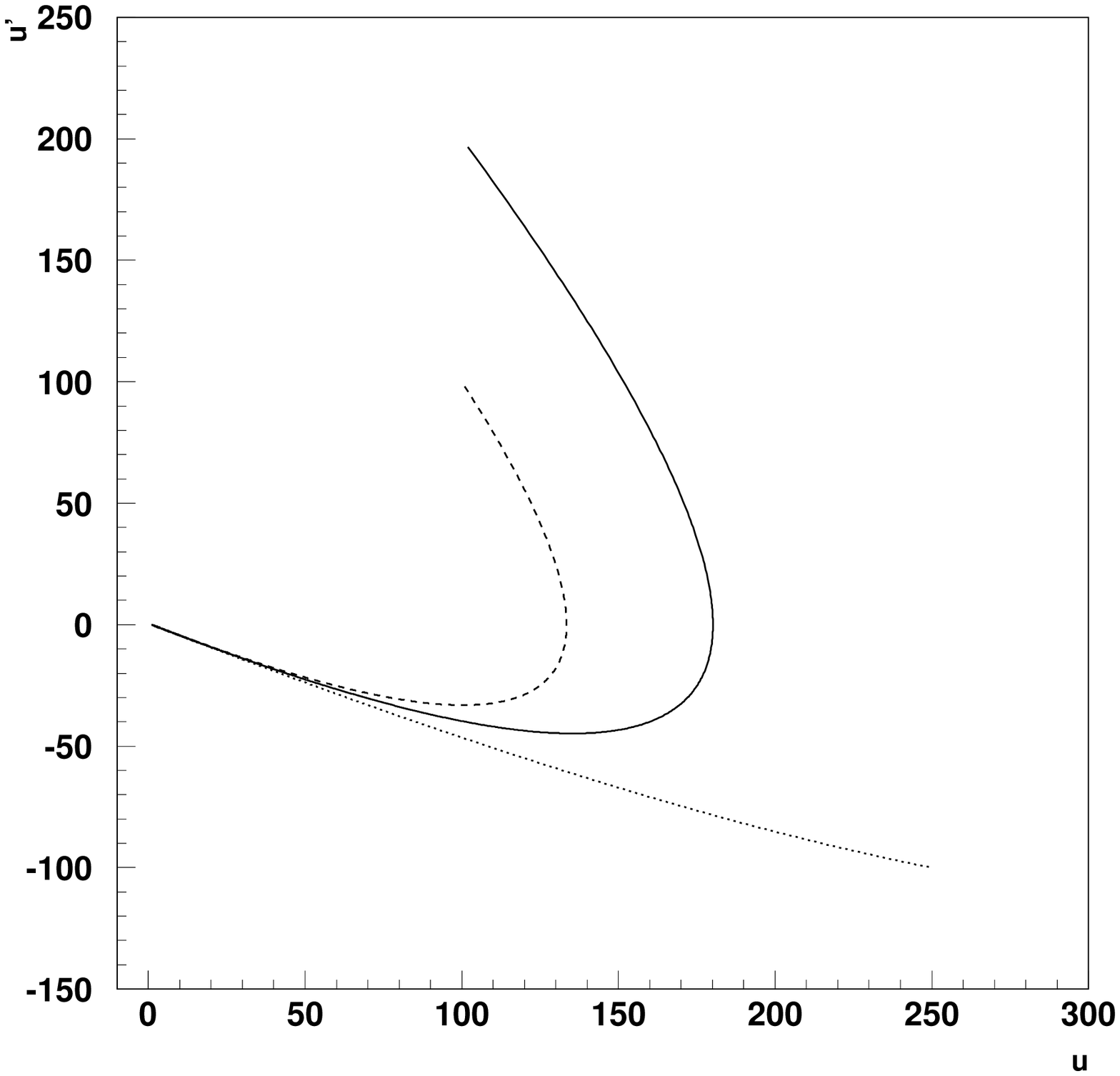}

 \caption{Numerical solutions of eq.(\ref{kg__umdivn}) are shown: on the
left side $u$ (black) and $u'$ (red) are plotted versus $\tau$, in
the case ($m = 4 \, , \, n = 2$) and converge to the stable
attractor solution $(u = 1 \, , \, u' = 0)$. Different curves
correspond to different initial conditions for $u$ and $u'$:
$(u_0, u'_0) = (100,200)$  (solid), $(100,100)$ (dashed),
$(250,-100)$ (dotted). The $\tau$ scale is logaritmic. On the
right side $u'$ vs $u$ is plotted for the same initial
conditions.} \label{figura1}
\end{center}
\end{figure}

The numerical solution of equation (\ref{kg__umdivn}), in full generality and
without linearization, is shown in 
fig.(\ref{figura1}) for a somewhat large set of initial conditions.
On the left side the time dependences of $u$ and $u'$ are shown, in the
case ($m = 4 \, , \, n = 2$). Different curves
correspond to different initial conditions for $u$ and $u'$ and they all converge
to the stable attractor solution $(u = 1 \, , \, u' = 0)$. On the right side it
is shown the $u'$ vs $u$ plot for the same choices of initial conditions of the
left hand side plot.

\subsection{Attractor behavior for $m = n$}
In the case $m = n$ the exact solution $\phi_e$ depends on time as in eq.(\ref{phi_meqn}) thus we have \be
\label{ffphi_mn_e} \frac{\rho_{mnr0}}{2a^3}\frac{F_{\phi}}{F}= \frac{A}{t^2} \left(
 \frac{6}{m} - 1 \right) \vv \ee where $A$ is the same as
  in expression ($\ref{phi_meqn}$). With the change of variable
 (\ref{tau_def}) the Klein Gordon equation becomes\footnote[3]{While writing the present
paper, the authors recognized that eq.(47) of Pettorino et al \cite{pettorino_baccigalupi_etal_2005} was obtained
under wrong assumptions. Namely, the value of $\phi_e$ found in the case $m \neq n$ was used in matter dominated
era (MDE) too, leading to wrong eigenvalues.}: \be
 \label{kg__umn}u'' + u' \left(\frac{2A}{\phi_e} + \frac{6-m}{m}
 \right)= \frac{A}{\phi_e}\frac{6-m}{m} (1-u) \pp \ee
Note that, unlike what happens in the previous case, the
coefficients in eq.(\ref{kg__umn}) depend on time through
$\phi_e(t)$. Nevertheless, $(u = 1 \,\, , \,\, u' = 0)$ still
behaves as a critical point for the equation and we are still
allowed to consider a generic perturbation $\delta u$ to the
critical point such that $u = 1 + \delta u$. Eq.(\ref{kg__umn})
then becomes: \be \label{kg_mun_pert} \delta u'' + \delta u'
\left(\frac{2A}{\phi_e} + \frac{6-m}{m} \right) =
-\frac{A}{\phi_e} \frac{6-m}{m} \delta u  \pp \ee However, we are
now dealing with a homogeneous differential equation of second
order in which the coefficients vary with time: \be \delta u'' +
P(\tau) \delta u' + Q(\tau) \delta u = 0 \vv \ee where \bea
\label{P_Q_def} P(\tau) = \frac{2A}{A\tau + \phi_*}+\frac{6-m}{m}
\\ Q(\tau) = \frac{A}{A\tau + \phi_*} \cdot \frac{6-m}{m} \pp \eea
In order to find the expression of the perturbation $\delta u$ we
consider the following change of variables: \be
\label{delta_z_def} \delta u = \delta z
e^{-\frac{1}{2}\int{P(\tau)d\tau}} \ee in terms of which
eq.(\ref{kg_mun_pert}) can be rewritten (if $\delta u \neq 0$) as:
\be \delta z'' + \left[ Q(\tau) -\frac{1}{2} P'(\tau)
-\frac{1}{4}P^2(\tau) \right]\delta z = 0 \pp \ee It is easy to
check that for our values of $P$ and $Q$ we get \be \delta z''
-\frac{(6-m)^2}{4m^2} \delta z = 0 \ee whose generic solution is
\be \delta z = c_1 + c_2 e^{\frac{(6-m)^2}{4m^2} \tau} \vv \ee
where $c_1$ and $c_2$ are arbitrary constants. Substituting this
expression in (\ref{delta_z_def}) we get the form of the
perturbation $\delta u$: \be \delta u = \frac{c_1 k}{\tau +
\frac{\phi_\ast}{A}} e^{-\frac{(6-m)}{2m} \tau} + \frac{c_2
k}{\tau + \frac{\phi_\ast}{A}} e^{\alpha \tau} \vv \ee where $k$
is a constant and $\alpha = \frac{3(2-m)(6-m)}{4m^2}$. It's then
immediate to see that the generic perturbation goes to zero for $2
< m < 6$, a range including both radiation $(m = 4)$ and matter
$(m = 3)$ dominated backgrounds, thus making $\phi_e$ a stable
attractor.

\begin{figure}
\begin{center}
   \includegraphics[width=7.5cm]{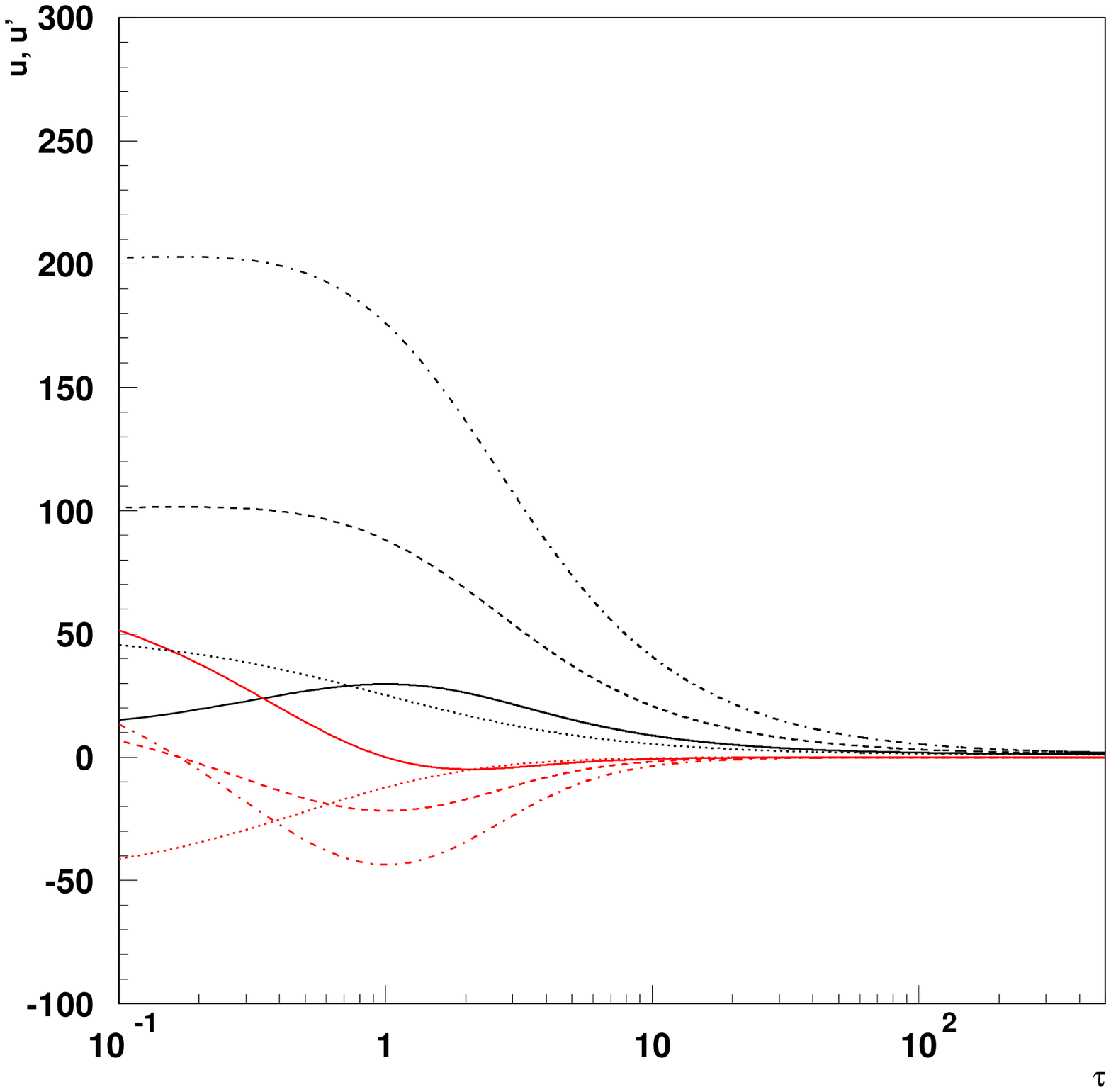}
   \includegraphics[width=7.5cm]{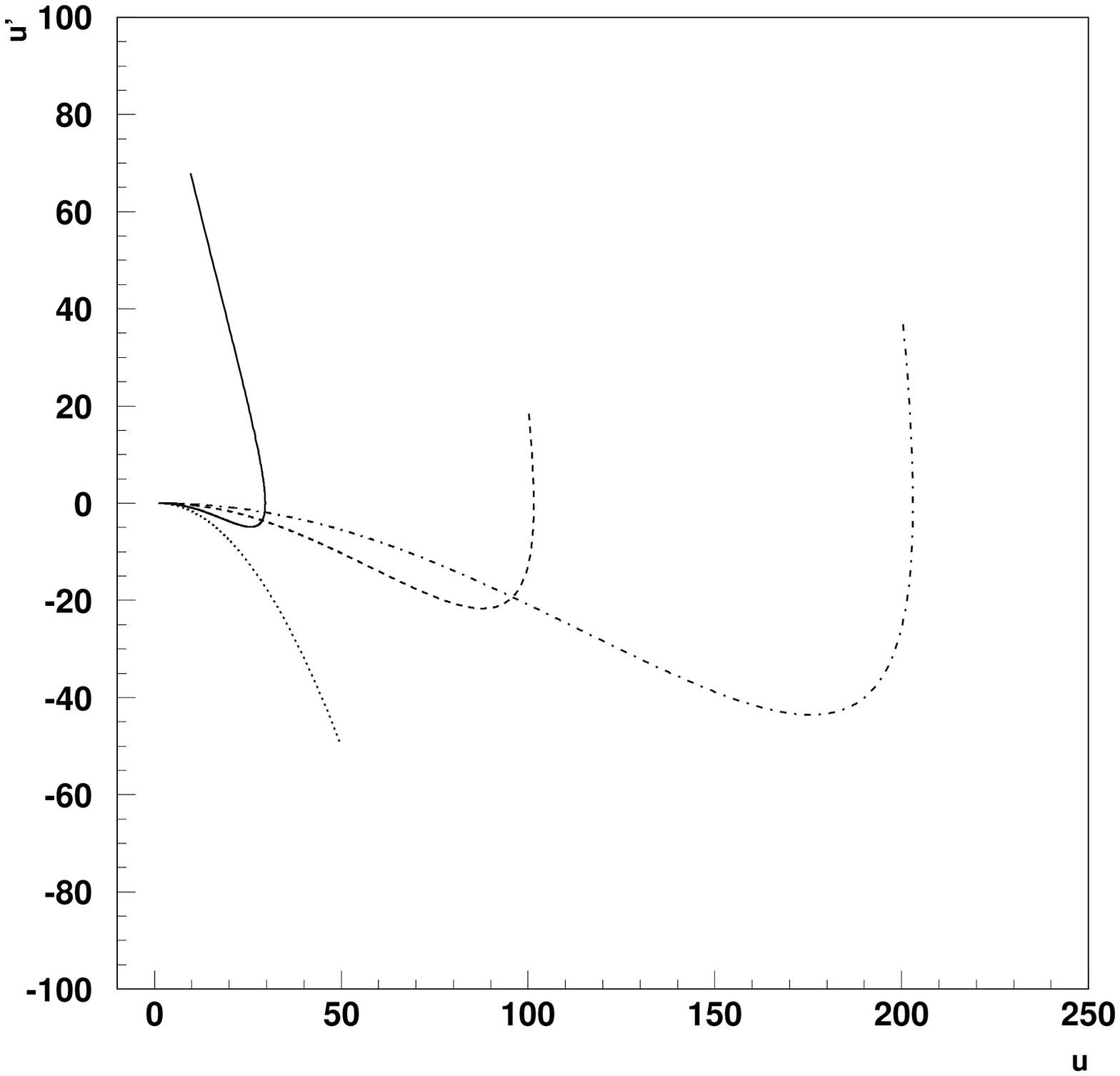}

\caption{Numerical solutions of eq.(\ref{kg__umn}) are shown: on
the left side $u$ (black) and $u'$ (red) are plotted versus
$\tau$, in the case ($m = 3 \, , \, n = 3$) and converge to the
stable attractor solution $(u = 1 \, , \, u' = 0)$. Different
curves correspond to different initial conditions for $u$ and
$u'$: $(u_0, u'_0) = (9,70)$ (solid), $(100,20)$ (dashed),
$(50,-50)$ (dotted), $(200,40)$ (dot-dashed). We have chosen the
test values $A = M_{Pl}$ and $\phi_0 = M_{Pl}$. The $\tau$ scale
is logaritmic. On the right side $u'$ vs $u$ is plotted for the
same initial conditions.} \label{fig_numerical_sol_mn}
\end{center}
\end{figure}

The numerical solution of equation (\ref{kg__umn}) is shown in
fig.(\ref{fig_numerical_sol_mn}) for a somewhat large set of initial conditions.
On the left side the time dependences of $u$ and $u'$ are shown, in the case
($m = 3 \, , \, n = 3$) and for the test values $A = M_{Pl}$ and $\phi_0 = M_{Pl}$:
different curves correspond to different initial conditions for $u$ and $u'$ and they
all converge to the stable attractor solution $(u = 1 \, , \, u' = 0)$. On the right side it
is shown the $u'$ vs $u$ plot for the same choices of initial conditions of the left hand side plot.


\section{Conclusions}
\label{conclusions} The analysis of the most recent data from
different cosmological probes such as Type Ia Supernovae, cosmic
microwave background, large scale structure, indicate that the
dark energy closely mimics a cosmological constant, and that the
crossing of the ``phantom line'', i.e. equation of state less than
-1 is currently allowed. While these results represent a severe
constraint for minimally-coupled quintessence models, a
non-minimal coupling has been shown able to encompass it, by
virtue of the enhanced dynamics imprinted at early time by the
coupling itself. However, there is not a priori a particularly
motivated form of this coupling, and the range of possibilities in
which it can be selected is unlimited. From the cosmological point
of view, however, it would be unphysical, as well as unpleasant,
to rely on models suffering of the fine tuning problem for the
initial configuration. Focusing on scalar-tensor theories as a
viable model to connect Dark Energy and gravity, we have thus
explored the possible choices of the coupling between a scalar
field and the Ricci scalar, in order to select those ones giving
rise to scaling solutions for the Klein Gordon equation, in the
form (\ref{rhophi}). Our analysis resulted in selecting three
classes of couplings, namely functions of the exponential form
(\ref{f_meqn3}), polynomial functions (\ref{f_mdivnB1}), and the
exponential of polynomial (\ref{f_mdivn}), depending on the
coefficient $m$ characterizing the scaling (\ref{rhofluid}) and on
the value of $\widetilde{B}$ on (\ref{C_mdivn}). \\
Our analysis is complete, in the sense that it recovers all the
possibilities to have a scaling behavior. Most importantly, it has
been found that these solutions actually possess attractor
properties, which is an extremely appealing feature in view of the
old fine-tuning problem of minimally-coupled quintessence models.
In particular, we have shown that all the scaling solutions of the
form (\ref{rhophi}) with $n<3$ and $4$ in the matter and radiation
dominated era, respectively, represent attractors. Clearly, this
enforces the case for extending the theory of gravity beyond
general relativity, and opens a window on the solution of
``initial values'' problem. Cosmological models, characterized by
the non-minimal couplings we have selected out in this papers,
will deserve a further investigation.

\ack We are pleased to thank G.Mangano and E.Linder for useful support, discussions and comments.

\end{document}